\documentclass[11pt, twocolumn]{article}
\usepackage{amsmath}
\usepackage{authblk}
\usepackage{graphicx}

\date{}

\begin{document}

\title{\huge Targeted Ads and/as Racial Discrimination: Exploring Trends in New York City Ads for College Scholarships}

\author[1]{Ho-Chun Herbert Chang}
\author[2]{Matt Bui}
\author[3]{Charlton McIlwain}
\affil[1]{Annenberg School for Communication and Journalism, University of Southern California}
\affil[2]{School of Information, University of Michigan}
\affil[3]{Steinhardt School for Media, Culture and Communication, New York University}

\maketitle

\begin{abstract}

\textbf{Abstract: }  This paper uses and recycles data from a third-party digital marketing firm, to explore how targeted ads contribute to larger systems of racial discrimination. Focusing on a case study of targeted ads for educational searches in New York City, it discusses data visualizations and mappings of trends in the advertisements’ targeted populations alongside U.S census data corresponding to these target zipcodes. We summarize and reflect on the results to consider how internet platforms systemically and differentially target advertising messages to users based on race; the tangible harms and risks that result from an internet traffic system designed to discriminate; and finally, novel approaches and frameworks for further auditing systems amid opaque, black-boxed processes forestalling transparency and accountability.
\newline

\end{abstract}

\section{Introduction}


\begin{quote}
    “While classifications are also parts of identificatory processes, their application is not primarily orientated toward particular individuals but toward particular types of individuals” (Gandy, 2021, p.6). 
\end{quote}

The 2016 fall-out of Cambridge Analytica ushered in greater scrutiny over the implications of Big Tech in everyday life. For one, the potential for discrimination and other potential harms and risks under targeted advertising have come under fire, with growing privacy debates and antitrust hearings (for example, see~\cite{gibson_us, reuters_2021}). Of note,~\cite{mikians2012detecting} and~\cite{hannak2014measuring} have documented cases of discriminatory pricing in e-commerce and travel, and crowd-sourcing in online advertising. Facebook, in particular, defended a civil rights lawsuit for targeting based on ``ethnic affinity''. This attribute was eventually renamed ``multicultural affinity" and the company promised to disallow ads related to housing, employment, and financial services to be targeted through this attribute~\cite{speicher2018potential}. Subsequently, articles on topics such as ``How Facebook's Ad Delivery Can Lead to Biased Outcomes" and ``Potential for Discrimination in Online Targeted Advertising" have become frequent topics of discussion in academic circles~\cite{ali2019discrimination, speicher2018potential}. Yet, despite increased activity in data ethics and fairness research, concerns are growing over the “ethics-washing” and flattening of critiques of technology’s social and political influence, especially as Tech companies
adopt---and arguably, co-opt---the language of important critiques of technology~\cite{le2020we}.

Drawing from a body of work including Cottom~\cite{cottom2017lower}, Gandy~\cite{gandy2021panoptic,gandy1993panoptic}, Benjamin~\cite{benjamin2019race}, and Noble~\cite{noble2018algorithms}, we examine online targeted ads as a technology-enabled form of social sorting: specifically, racial discrimination. This language and framework is in conversation with data ethics and fairness discourse as a new but important remediation: to draw attention to structural (i.e., meso- and macro-level) inequities; and to move away from individualizing issues of data ethics and fairness, largely under the guise of “de-biasing". Moreover, we draw from recent algorithmic audit studies to empirically test for---and identify instances of---discrimination, while articulating potential next steps and novel methodologies for auditing discriminatory systems such as targeted advertising. This is all done to highlight the systemic and structural dimensions of digital harms and risks that motivate urgent calls for accountability interventions. 


\section{Background Literature}
\subsection{Technology as Enablers of Racial Discrimination}

Within the foreword of the second edition of \textit{The Panoptic Sort}, Gandy (2021) revisits his seminal work, originally published in 1993~\cite{gandy2021panoptic,gandy1993panoptic}. \textit{The Panoptic Sort} (1993) articulated and described the dangers of the growing accrual of individual personal data, particularly for social and economic purposes. Using marketing and employment as key examples, among others, one of his Gandy’s key arguments was that advanced computation and networked information systems both enable and amplify previous forms of racial and social sorting, as large volumes of personal data are used to sort, categorize, and “optimally” distinguish---discriminate---between individuals. 

Three decades later, this work still explicates prescient and persistent critiques regarding the oft-under-examined nature of contemporary commercial digital systems, especially targeted ads. Gandy writes about panopticism, connecting it with the social engineering of urbanization: ``The insanity of the urban core reflects a hopelessness that is reproduced by the operation of the panoptic sort--a discriminatory technology that selects out and rewards self-identification as deviant and dysfunctional and increases the sharpening of distinctions that are then reified and institutionalized. Panopticism identifies, breeds, cultivates, and reproduces failures" (Gandy, 2021, p. 258). Indeed, Gandy’s work on surveillant data-driven systems connects technologies with racially discriminatory processes, calling attention to classifications and power hierarchies---that is, processes of discrimination---reified and reproduced by technology within everyday urban life (see also~\cite{gandy1995s}).  

Benjamin~\cite{benjamin2019race} echoes these concerns regarding the digital and data-enabled forms of racial sorting and inequality, proffered by various contemporary modes of technology design and development. In particular, Benjamin highlights how both techno-benevolence and techno-ignorance---i.e., technology designers’ good intentions and their lack of critical understandings of Others' lived experiences---enable and reproduce inequality through technology. Meanwhile, Noble~\cite{noble2018algorithms} critiques the commercial outsourcing and commodification---through Google Search Results---of reference materials, resources, and information, drawing attention to the advertising revenue schemes that uphold racism, sexism, and other forms of discrimination, by and through online search tools~\cite{noble2018algorithms}. In short, Gandy, Benjamin, and Noble, drawing from commitments to critical scholarship that center marginalized communities and their concerns, articulate the connections between technology and racial discrimination, providing analytical lenses for describing, examining, and understanding how technologies are, in truth, built and designed to \textit{discriminate}; and how these discriminatory tools are not bugs---but rather \textit{features}---of such systems, profiteering from categorization and targeting processes.

Thus, discrimination---vis-a-vis fairness---is an important yet novel frame for emerging conversations about the social implications, and ethics, of data and AI. Namely, algorithmic audits have emerged as a growing field and topic of intervention in response to data harms and risks. Audits are often sector specific, such as job advertising, and specify protected categories, such as race and gender. Bias is characterized as skew, and is then interpreted in accordance to some criteria, for instance whether skew is explainable by differences in qualifications from other factors~\cite{imana2021auditing}, and has been shown in gender~\cite{datta2015automated} and race~\cite{sweeney2013discrimination}.

Due to these outcomes, many studies also focus on the algorithms themselves, specifically hoping to provide explainability and thus identify sites for programmatic intervention, for instance, as demanded by the ``right to explanation" in service of the EU General Data Protection Regulation~\cite{wachter2017counterfactual}. Proposed solutions have included simulation and probabilistic solutions, such as relying on counterfactual explanations to generate sets of recommendable actions~\cite{karimi2021algorithmic}, using the machinery of probabilistic graphs. Google, in collaboration with Inioluwa Raji, has also envisioned means of an end-to-end framework for internal audits, primarily focused on the various stages of an algorithms production and deployment~\cite{raji2020closing}. 

Yet, interventions have generally focused on “computational” tools for “de-biasing”, lacking a concerted engagement with the social and historical dynamics of issues of data ethics and fairness~\cite{dencik2019exploring}. Hoffmann (2019) calls attention to this, by critiquing the limits of “anti-discriminiation” discourses within data ethics conversations~\cite{hoffmann2019fairness}. Meanwhile, Dencik~\cite{dencik2019exploring} foregrounds ''data justice" as a novel distinction from ''fairness", namely in its interconnectedness with social and economic justice; whereas~\cite{milan2016alternative} discusses ``data activism" as connecting data with the vantage point of users and impacted communities. Put simply, whereas many see computational tools for de-biasing as a fruitful endeavor, others scholars call attention to the language and discourse around issues of data ethics as an equally important area for intervention, often critiquing the fetishization---and, at times, epistemological arrogance and assumptions---of advanced computational tools for ``fairer" machine learning; and a general lack of interrogation regarding who benefits from the continued presence and expansion of such systems. 

\subsection{Online Ads, Data Access, and Zipcodes as Racial Proxies}	
As noted, a growing area of concern related to data ethics is the topic of targeted ads. Moreover, due to its status as a protected trait, race served as an important area of focus for the regulation and sanctioning of Facebook’s ad library~\cite{speicher2018potential}. 

Yet, as Pasquale (2015) explains, the online ecosystem is often a commercially-driven system protected by its “black boxes”---that is, its opaque and hard-to-access troves of data, often with restricted access for internal parties and actors. Thus, amid growing concerns and scrutiny for contributing to racial discrimination, rather than providing data for auditing, data platforms and tech companies have attempted to further hide or limit access to these data points, opting for opacity as a tactic to avoid regulation and sanctions. For this reason, in pursuit of preserving "alternative" methods of data collection and access, Sandvig and colleagues~\cite{sandvig_2019} challenged \textit{The Computer Fraud and Abuse Act}, arguing for the need for the protection of various algorithmic audits employing data scraping methods, especially with an eye toward addressing issues of racial discrimination.

Considering these debates and a general lack of evidence and data access for algorithmic audits without potentially compromising relationships or agreements (see~\cite{pasquale2015black, kerr_2020}, for example), we propose zip codes as another potentially promising, and novel interlocutor for important racial demographic data for algorithmic audit studies. In short, zipcodes (and their racial compositions) are often accurate predictors of life expectancy~\cite{spoer2019census}. Thus, using readily available census datasets to model racial and ethnic demographics (alongside other demographic variables) could supplement and empower algorithmic audit studies investigating issues of racial bias and discrimination in vital ways. (Admittedly, some companies have already begun to restrict access to geolocation data, through which racial and ethnic categories can be modeled and inferred.)

\subsection{Online Education and Discrimination}
It is then useful to be precise about the substrates of discrimination. As zipcodes have been traditionally used a significant feature for targeted advertising, the dangers here (at least for the contemporary legal environment) are not necessarily infringement upon the law. Here, \textit{predatory inclusion} is a  useful concept that characterize how zipcode-based advertisements may contribute to aggravating pre-existing discriminatory systems~\cite{mcmillan2020platform}.

Predatory inclusion is the “logic, organization, and technique of including marginalized consumer-citizens” into extractive schemes~\cite{cottom2017lower,mcmillan2020platform}. Education here is a prime example. For instance, McMillam Cottom articulates how African American woman may be targeted by online college degrees, then convinced to take on student loans that have a higher risk of default and negative amortization. Rather than a direct violation of race-specific laws, the dangers presented by predatory inclusion demonstrates an asymmetry of risk resulting from targeted advertisements. Thus, predatory inclusion and its risks are deeply related to finance~\cite{nopper2010colorblind, charron2020racialized}.

Second, higher education’s role in reinforcing inequality has been a long-standing topic of research, in part due to its equally long-standing history within the United States. Socioeconomic status~\cite{kincheloe2007cutting}, neighborhood effects of universities~\cite{cordes2016neighbourhood, jencks1990social}, and the private/public divide~\cite{newfield2018great} all intersect significantly with race. Therefore, the specific type of university and the demographics they target can potentially reveal whether these extant inequalities are being perpetuated, reinforced, or ameliorated by algorithmic systems.

As such, while companies and universities targeting different communities based on demographics is by no means novel, technology’s role in further enabling this discrimination is of great importance. For instance, the \textit{Harvard Business Review} showed individuals in wealthier areas respond more strongly to e-commerce discounts. This paradigm where higher income individuals have access to lower prices is one way technology uniquely enables discrimination~\cite{miller2019targeted}. The implementation of control theory in marketing in effect preserves the "equilibrium": in other words, marketing strategies proffer models that move further towards reifying pre-existing inequalities~\cite{couldry2019data}.

\subsection{Research Questions}
In summary, marketing is inherently tied to its target audiences and targeted advertising has been justified by, designed for, and predicated upon the need for personalization based on an individual’s characteristics. Yet, amid a broader moment of reflection and scrutiny over technology-enabled injustices and discrimination, this paper questions whether all targeted ads---and the companies buying ad space---should presuppose their target audiences, especially within the realm of educational opportunities. Put simply, considering how higher education is often construed as a powerful tool for social and economic mobility, should there be targeted audiences for ads for educational opportunities? Moreover, considering Taylor’s~\cite{taylor2019race}  and Mcmillan Cottom’s~\cite{cottom2017lower} work on predatory educational “opportunities” predicated on inclusion, who is more likely predisposed to such schemes? Finally, how do we track, measure, and understand more deeply the loss of opportunities---that is, the harms and risks of technology-enabled racial discrimination---within and across racial categories? 

Thus, this paper thus has two primary goals. First, using zip-codes (and zip-codes alone) as the basis of our analysis, it seeks to analyze the differences in ad coverage across multiple market sectors. The goal of investigating multiple sectors is to first illustrate how different products can diverge greatly, the types of targeting that occurs, and the primary actors in each sector responsible for biased behavior. Second, it seeks to build a context-based model that allows us to assess whether evidence merits the classification of racial discrimination: namely, we propose to use content analysis (in future studies) to disentangle what is being marketed to different racial markets.

In all, this study touches on important and emerging concerns regarding platform accountability, regulation, and ethics. We anticipate our results will be relevant to discussions about tech policy, and tech and society debates more generally. Our guiding questions were: 
\begin{itemize}
    \item \textit{How do online ads discriminate within and across communities?}
    \item \textit{Why do we need to deconstruct how targeted ads are, by nature, racially discriminatory?}
\end{itemize}

To examine them, our research questions for this study were as follows: 
\begin{enumerate}
\setlength\itemsep{0em}
    \item Where are the top domains targeting their ads (in terms of zip code)?
    \item Which zip codes are targeted for the best and worst (employment and housing) opportunities? (Is there a bias in the “hotspots” for online ads?)
    \item To what degree do these distributions of online ads reflect current and historical racial-spatial inequalities (i.e., segregation)? (Are race and/or class strong predictors of ad targeting?) 
\end{enumerate}

\section{Methods}

\begin{table*}[tb]
\small
\begin{tabular}{lllll}
\textbf{Keyword}        & \textbf{Volume} & \textbf{CPC (USD)} & \textbf{\begin{tabular}[c]{@{}l@{}}Competitive \\ Density\end{tabular}} & \textbf{SERP Features}                         \\
covid-19                & 3,350,000         & 0                  & 0                            & Knowledge Panel, Top Stories, Site Links       \\
black lives matter      & 165,000          & 1.25               & 0.05                         & Knowledge Panel, Top Stories, Site Links       \\
houses for rent near me & 823,000          & 0.33               & 0.46                         & Image Pack, Adwords Bottom                     \\
college scholarships    & 40,500           & 1.9                & 0.63                         & Top Stories, Site Links, Adwords Top           \\
online degree programs  & 4,400            & 27.14              & 0.94                         & Site Links, Adwords Bottom, People Also Ask   
\end{tabular}
\caption{Keywords for market sector analysis.} \label{tab:market-sector-keywords}
\end{table*}

\subsection{Data}
Our data was procured through SEMRush's \textit{Competitor Discovery} platform. By inputting a zipcode and keyword, we tracked the top 80 to 120 domains that vie for each ad keyword, including their \textit{rank} (within Google Search Results), \textit{relative visibility}, and \textit{estimated traffic}. As an example, our primary interests were in the housing and online education sectors, so we first inputted our sectors of interest (i.e., education) as a seed keyword, and then picked the top keyword by search volume. Table~\ref{tab:market-sector-keywords} shows the keywords by sector or topic. For the purpose of this study, we focus on education, in particular \textit{college scholarships}. 
Together, this yielded 248,884 url-zipcode pairs for New York and 191,697 url-zipcode pairs for Los Angeles.

\subsection{Census Covariate Aggregation}
We extracted census data for all zip codes for Los Angeles and New York City, using the official ACS API~\footnote{https://github.com/jtleider/censusdata}. The primary covariates were as follows:
    \begin{itemize}
    \setlength\itemsep{0em}
        \item Total Race
        \item White alone
        \item Black alone
        \item American Indian alone
        \item Asian alone
        \item Native Hawaiian Pacific Islander alone
    \end{itemize}


\subsection{Network Construction}
To recap the structure of our data, each zip-code was associated with a list of: a) domains that vie for a keyword; and b) the estimated traffic for that domain. Let $i$ denote each unique domain and $x_{i,a}$ the traffic of domain $i$ in zip-code $a$. The similarity between zip-codes $a$ and $b$ can then be written explicitly as:

\begin{equation}
    dist(a,b) = \sum_{\forall i \in I} ( x_{i,a} - x_{i,b} ) ^2
\end{equation}
\begin{equation}
    sim(a,b) = \frac{1}{dist(a,b) + 1}
\end{equation}
where $I$ is the union of all domains available in the data set. In short, we first computed the Euclidean distance between two zip-codes, based on domain-specific pairings, and then defined the similarity measure as its multiplicative inverse. Note, we added 1 to ensure domains with a distance of 0 could be computed, while preserving order. A similarity measure of 1 indicates the exact same domains and relative traffic levels.

For every keyword, we constructed a network $G(V,E)$. Each zip-code is a node, and the pairwise similarity scores are the weight. We then visualized differences: a) at the keyword level and b) city-level (Los Angeles and New York). Additionally, we visualized the distribution of similarity weights to understand the variance in marketing for each keyword. Upon constructing the network, we then applied community detection techniques to identify clusters of similar zip-codes. For graphs with a large variance of similarity scores (edge weights), we expected there to be clear clusters. In contrast, a graph with uniform edge weights of 1 indicated that the domains and visibility were the same across all zip-codes.

\subsection{Domain-Level Analysis}
With census data, we model the demographics and racial population for every zipcode. 
We sought to evaluate the biases each domain produced in terms of identifying their target audience(s). Due to missing and incomplete data for other racial/ethnic categories (e.g., Hispanic and Latinx populations), our analyses of New York City focused on the following racial groups: \textit{White}, \textit{Asian}, and \textit{Black}. However, the total population will include all racial categories.
Formally, we formulate the net demographics on each domain. Let $\vec{v}_z$ be the population of each racial category on each zipcode $z$:
$$
\vec{v}_z = 
\begin{bmatrix}
p_{White} \\ p_{Asian} \\ p_{Black}
\end{bmatrix}
$$
Next, let $\theta \in \mathbf{N}$ denote a cut-off for the number of zipcodes to consider. Here we set $\theta=20$, given that we include 202 unique zipcodes in our greater New York City region and this would correspond to 10\%. Let $R(d,z) = i$ denote the ranking function for domain $d$ in zipcode $z$, which generates the ordered set of visibility scores for domain $d$. Thus, the set of all proportions bound under our cut-off $\theta$ is denoted:

\begin{equation}
    V(\theta, d) = \Big\{ \vec{v}_z \quad | \quad \forall \vec{v}_z 
    \text{  s.t.  } R(d,z) > \theta \Big\}
\end{equation}
In other words, $V(\theta, d)$ contains the demographics for the top $\theta$ zipcodes that domain $d$ bids on. We then calculate the domain's target population as follows.
\begin{equation}
\begin{aligned}
      U(d) &= \sum_{ \vec{v} \in V(\theta, d)} \vec{v}   \\
      P(d) &=  \frac{U(d)}{ ||U\*(d)|| }
\end{aligned})
\end{equation}
$P(d)$ here captures the net racial demographic for a given domain $d$. 
In plain language, $U(d)$ denotes the total Black, Asian, and White individuals in domain $d$'s top zip codes, in a $3\times 1$ vector. $P(d)$ is the total number of individuals for a given racial demographic, divided by the total population across these zipcodes. Note, $||U\*(d)||$ is the total population including the redacted values from other racial categories.
We then compared it with the city-average to see if domains target zipcodes on the basis of race: i.e., if they tended to target areas predominantly Asian, Black, and/or White. For this value, there are two interpretations. The first is a literal, that a specific area yields the most exposure for a given domain. This disregards the intentions of the domain (and its underlying business). The second interpretation is that it signifies where a domain spends more of its resources bidding for exposure in these target zipcodes. 

\begin{figure}[!htb]
    \centering
    \includegraphics[width=1.0\linewidth]{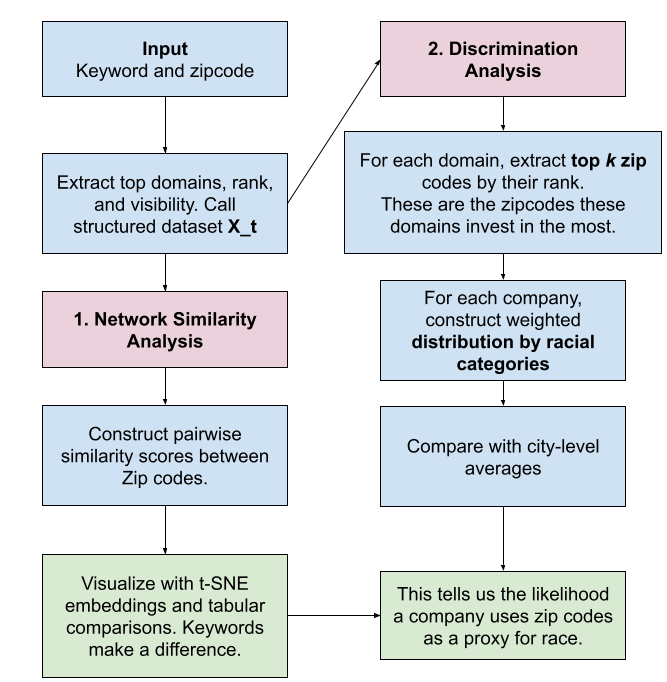}
    \caption{Overview of the methodological design.}
    \label{fig:method}
\end{figure}

While the similarity analysis establishes differences exist on the macro scale, how domains differ is in need of further investigation. To conclude, Figure 1 illustrates the method logical pipeline and progression of our study.

\begin{figure*}[!hbt]
    \centering
    \includegraphics[width=1.0\linewidth]{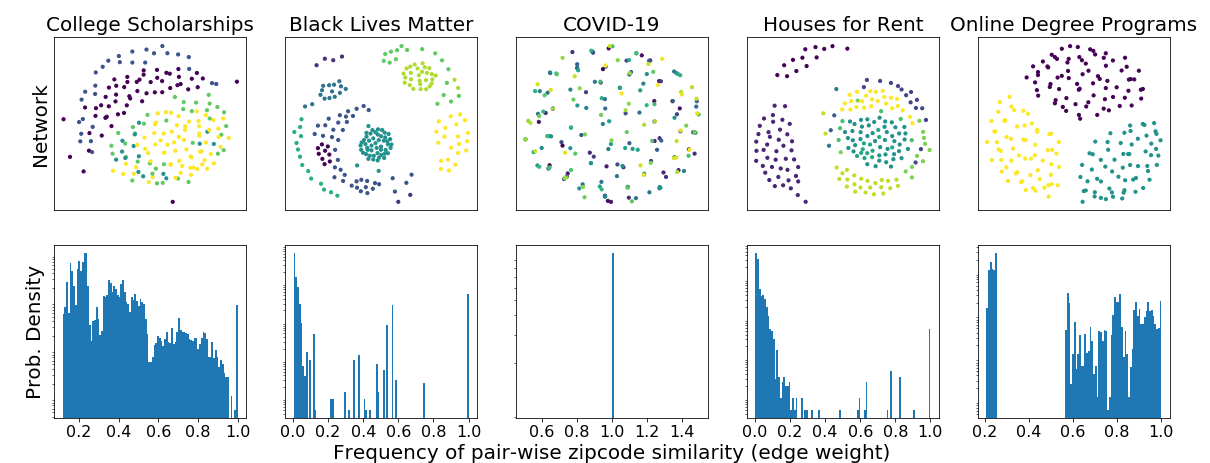}
    \caption{Overview of the methods used in this study.}
    \label{fig:similarity}
\end{figure*}

\section{Results}
\subsection{Overview of Differences across Keywords}

Our first research question is meant to establish whether we find differences in domains' targeting strategies, in terms of the similarity of zipcodes they reach (across keywords). Figure~\ref{fig:similarity} shows the network topology of all the keywords from Table 1, which are then colored using the Leiden community detection algorithm (which boasts better runtime and accuracy when compared to the Louvain algorithm)~\cite{traag2019louvain}. The zipcodes are then plotted using a forced-based layout.

We observe some immediate differences. For one, clear clusters can be observed with the keywords \textit{Black Lives Matter}, \textit{Houses for Rent}, and \textit{Online Degree Problems}. On the other hand, keywords such as \textit{COVID-19} seem to be more well-mixed (i.e., little to no clustering). This can be explained by observing the distribution of pairwise similarity weights, shown in the second row. COVID-19, for instance, shows only one value of similarity weight---with w = 1. This is because across all zip codes, ads about the pandemic---primarily from the CDC and state of New York---are congruent and targeting all zipcodes equally. That is, no differences nor discrimination exists; and indeed, there is no clear network clustering for COVID-19 in Figure 2. On the other hand, keywords such as \textit{College Scholarships} and ones that evoke race directly demonstrate a spectrum of similarity weights, and we observe a corresponding “purity” of clustering in Figure 2.

Together, these two figures show that, at the macro-scale, there are differences across keywords. Network clustering is more prevalent in cases of racially charged keywords, whereas COVID-19, a national-level issue, exhibits no differences across zip codes.

\subsection{Keyword Case Study: College Scholarships}
Our next step is to establish how these targeting strategies are different. We focus our attention on the education sector, with the keyword \textit{College Scholarship}. As a brief reminder, for every domain, we consider their top $\theta$ zip codes for which they compete for, in terms of visibility rankings.  Table~\ref{tab:top_domains_race} shows the top domains based on our three racial categories for New York City. 

A few observations can be made. First, the top contenders for the keyword consist of universities, external scholarship providers such as \textit{coca-colascholarsfoundation.org}, and general tertiary education websites such as \textit{studentscholarships.org} and \textit{finad.org} which aggregate scholarship information. This includes government-sponsored websites such as \textit{studentaid.gov}.

\begin{table*}[bt]
\begin{tabular}{llll}
\textbf{Rank} & \textbf{White}                  & \textbf{Black}            & \textbf{Asian}     \\
1             & landmark.edu                    & collegeboard.org          & myscholly.com      \\
2             & studentscholarships.org         & phoenixpubliclibrary.org  & hope.edu           \\
3             & firstinspires.org               & fastweb.com               & contracosta.edu    \\
4             & collegesofdistinction.com       & cuny.edu                  & wvu.edu            \\
5             & coca-colascholarsfoundation.org & collegescholarships.com   & macomb.edu         \\
6             & compostfoundation.org           & unigo.com                 & gocolumbia.edu     \\
7             & mometrix.com                    & jumpstart-scholarship.net & alpenacc.edu       \\
8             & pinterest.com                   & collegegreenlight.com     & palmbeachstate.edu \\
9             & ed.gov                          & sfcollege.edu             & ccis.edu           \\
10            & spcollege.edu                   & discover.com              & bestcolleges.com  \\
11            & dickinson.edu                   & scholarships.com          & ncc.edu            \\
12            & nitrocollege.com                & uncf.org                  & scholarshipowl.com \\
13            & affordablecolleges.com          & studentaid.gov            & schoolcraft.edu    \\
14            & columbia.edu                    & wgu.edu                   & templejc.edu"      \\
15            & foxnews.com                     & mdc.edu"                  & spelman.edu        \\
16            & finaid.org                      & evergreen.edu             & central.edu        \\
17            & meredith.edu                    & meredith.edu              & ccp.edu            \\
18            & evergreen.edu                   & triton.edu                & tallo.com          \\
19            & pct.edu                         & collegescholarships.org   & phoenix.edu        \\
20            & niche.com                       & salliemae.com             & gordon.edu         
\end{tabular}
\caption{ Top domains for each racial category (White, Black, and Asian) for the city of New York} \label{tab:top_domains_race}
\end{table*}

\begin{figure}
    \centering
    \includegraphics[width=0.9\linewidth]{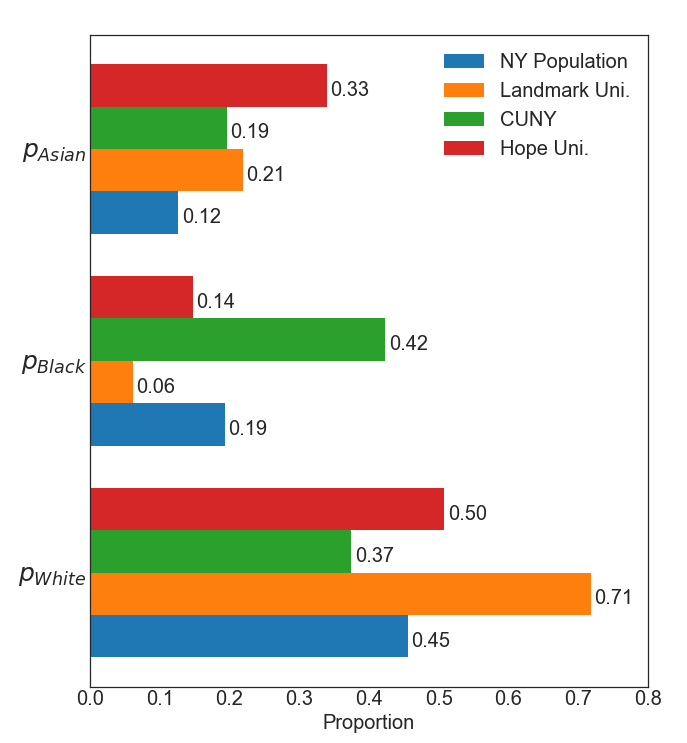}
    \caption{Relative demographics of served ads by university. Landmark University}
    \label{fig:top_prop_NYC}
\end{figure}

We turn our attention to the university level, considering the different “.edu” domains present in the dataset.  Figure~\ref{fig:top_prop_NYC} shows the domains that generate the greatest biases as a result of their bidding strategy, based on their relative target demographic. To recap, the relative demographics--- normalized on White, Black, and Asian racial categories--- better compares across these three groups.

First, note the relative demographics for New York City at large is as follows: 12\% Asian, 19\%Black, and 45\%White. Landmark University (in orange), which is the top domain in terms of targeting White audiences, serves 71\% of its ads to White zipcodes, 6\% to Black zipcodes, and 21\% to Asian zipcodes. Meanwhiile, CUNY (in green) serves ads mostly to predominantly Black zipcodes (42\%); and a marketing proportion to White and Asian demographics at 37\% and 19\%, respectively.  In this case, Black audiences encounter ads at a rate of more than 2 times the expected rate, based on the city average. Finally, Hope University (in red) serves its ads accordingly: 33\% to Asian zipcodes, 50\% to White zipcodes, and 14\% to Black zipcodes; this rate for predominantly Asian zipcodes is almost 3 times the city average.

It is also helpful to see where universities' ads pull away from, given higher rates of targeting in specific racial demographics. For Hope University, an increase in the Asian demographic took away ads from Black zipcodes. The White population also saw a slight increase, from 0.45 to 0.50.  CUNY, however, derives its increased Black audience from a drop in its White audiences. Lastly, Landmark University generates much of its increased White audiences (0.71 against the city-wide average of 0.45) at the cost of a diminished Black audience (0.06 against the city-wide average of 0.19). The Asian marketing efforts remain largely level.

\begin{figure}
    \centering
    \includegraphics[width=1.0\linewidth]{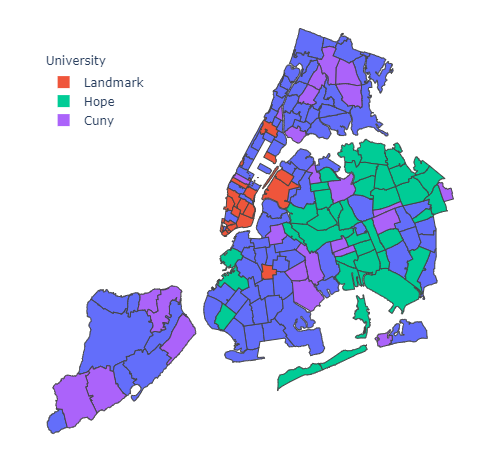}
    \caption{The primary bidding targets of the top three domains by divergence from city-wide racial demographics.}
    \label{fig:map}
\end{figure}

We interpret these results relative to the city-wide averages. The different rates of targeting may be from a variety of reasons, due to allocation of marketing budget to deliberate, race-based choices. In the best case scenario, we observe distinctions in who different universities bid for. Additionally, these differences in levels may certainly be attributed to characteristics of specific zip codes, such as income. To study the geospatial dimension, Figure~\ref{fig:map} shows a map of New York, its five boroughs and the top 20 zip codes that Hope, CUNY, and Landmark dedicate their bidding efforts. 

Landmark's (orange) audience can be seen as predominantly around the Manhattan area and parts of Brooklyn. In contrast, Hope University's dominant audience is found in the Queens area and Long Island, with some bidding in South Brooklyn. CUNY, in contrast, bids the most across all five boroughs, with top bids in Staten Island, Manhattan, Brooklyn, and notably the Bronx, which is the largest departure compared to the other two schools.

\begin{figure}
    \centering
    \includegraphics[width=1.0\linewidth]{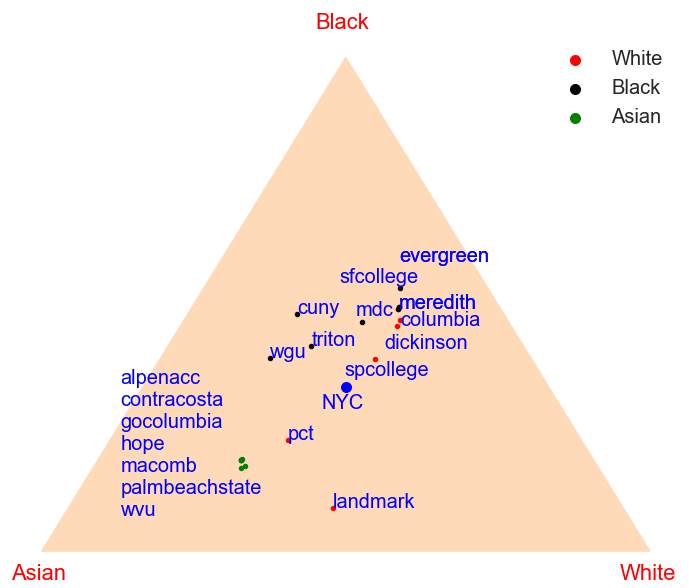}
    \caption{Ternary diagram of top universities by racial category, centered by the New York City average. Proximity to one corner indicates an ad-bidding bias toward that racial demographic. }
    \label{fig:ternary}
\end{figure}

These three schools are singular examples. To generalize, we plotted all  of the schools found within the top twenty domains in Figure~\ref{fig:ternary}. Ternary diagrams are useful for analyzing interactions across three categories~\cite{chang2021elitism}.
This ternary diagram indicates the relative bias of each school, based on the city-average. The city average is found in the center, which corresponds to  the vector $(0.45, 0.19,0.12)$, or the proportions of the three racial categories. A point found on the bottom left corner indicates the domain bids for ads for an audience that is 100\% Asian comprised: a point at the top for the Black demographic; and at the bottom right for White audiences. Furthermore, points lying opposite of the corner (in the middle of the opposite edge), indicate a bias toward the two other categories. For instance, WGU is split evenly between the Black and Asian demographics, at the cost of the White demographic. 

In combination with Table~\ref{tab:top_domains_race}, there are a few obvious observations. First, the top domains bidding for the Asian demographic are larger in number (15) compared to the number of universities vying for predominantly White and Black zipcodes (5 and 6 respectively). Furthermore, most of the universities that targeted predominantly Asian zipcodes are clustered together. In conjunction with Figure~\ref{fig:map}, this indicates these schools likely bid on the same zip codes, particularly in Queens. 

In contrast, schools that prioritize other demographics are more spread out. Landmark University, for instance, seems particularly deviant from other domains bidding on mostly White zipcodes, i.e., the others bid mostly on Black and White zipcodes at the cost of Asian zipcodes. Given that the Black points lie opposite of the White corner, this indicates Landmark University strays away from predominantly Black zipcodes without compromising their targeting of Asian zipcodes. 

\section{Conclusion}
In this paper, we began by reviewing extant work on technology-enabled discrimination to inform our novel approach to measuring and identifying biases within targeted ads, especially in terms of their differential target audiences and priorities. Specifically, Benjamin, Gandy, and Noble highlight the structural inequities that often give rise to---and reify and institutionalize---racial hierarchies through technology. Moreover, the site of discrimination can arise in two ways---\textit{predatory inclusion} in accordance to Cottom~\cite{cottom2017lower} and these algorithms as ways of reinforcing pre-existing inequalities within higher education. We now attempt to re-interpret, refine, and connect our results in light of these concepts about the racial and social inequalities coded and encoded within technology; and broader systems of discrimination through technology.

First, our results focused on the keyword \textit{College Scholarships} as a case study for how we might analyze and audit discriminatory ad mechanisms in the future. We showed that there are stark divergences in how different university domains bid for a given keyword. For example, more universities bid for predominantly Asian demographics, and bidding is shown to be more geographically localized (around Queens). Meanwhile, others---whether intentionally or unintentionally---strategize and target predominantly White and/or Black areas. In contrast, the targeting for \textit{COVID-19} as a search query showed no differences in exposure to targeted ads (and domains).

These results are important considering equal access and opportunity protections within education. Given past research that documents how education can both overcome and re-instantiate social and racial inequality, it is important to further interrogate how targeted advertising---and commercial digital systems, more broadly---reproduce and exacerbate these inequities. In short, rather than attempting to ``de-bias" and subsume algorithmic models and audits into idealized worldviews and notions of fairness, discrimination as a frame highlights the social construction of technologies: they have often, and will likely continue to, discriminate, reproducing new and old forms of social and racial sorting within communities and society. Yet, as Gandy (2021, 1993) highlights, the identification and documentation of these issues is vital work for critical scholars, especially in efforts to contest and take power away from dominant imaginaries of technology--and rather, to assert new modes and models for truly fairer and more just technologies. 

Moreover, while it is important to consider how and why universities are marketing \textit{College Scholarship} in racially differential ways, these scholarships are often individually-awarded. Therefore, a stronger examination of structural and community-level inequities--racially discriminatory targeted advertising---might be seen through analyzing another keyword in our dataset, \textit{Online Degree Programs}. Indeed, our preliminary and exploratory analyses of this keyword demonstrate elite and Ivy League universities as prioritizing predominantly White zipcodes.

Considering zipcodes as inextricably tied to racial and demographic data---that is, thinking about zipcodes as better predictors of socioeconomic outcomes than advanced machine learning models---we propose zipcodes offer an underutilized view into facilitating algorithmic audits for identifying racial discrimination, given general lack of access and uncompromised data access for these studies. There is a reason for this strong correlation, and it is deeply tied to legacies of racial oppression that shape the zipcode as a highly informative data point and predictor for an individual's quality of life and livelihood.

More important than providing metrics for  biases and divergences, future research should continue to draw attention, and attend, to the structures and systems in which these processes take place. As such, we will be doing comparative work moving forward. We will draw cross-city comparisons (comparing New York and Los Angeles) and across keywords. Second, we will also conduct regression analyses to quantitatively specify the impact of other census variables--- in particular income and education level. Lastly, using an associated corpus of the text ads delivered by the domains, we will implement content analysis to specify how semantic content relates to disparities and discrimination in ad delivery. This future work will work toward contributing toward understandings of, and contraventions and interventions against--rather than ignoring and naturalizing--reinforcements and reproductions of racially discriminatory processes and racial hierarchies through technologies. 

\section{Acknowledgements} 
The authors would like to thank the Democracy Fund and the Center for Critical Race and Digital Studies at New York University for their ongoing support of this project.

\bibliographystyle{ieeetr}
\bibliography{sample}

\end{document}